\documentclass[letter]{aa} 

\usepackage{hyperref}
\usepackage{ csquotes}

\usepackage[utf8]{inputenc}
\usepackage{xcolor}
\usepackage{amssymb}
\usepackage[small,labelfont=bf]{caption}
\usepackage{graphicx}
\usepackage{hyperref}
\hypersetup{
	colorlinks,
	linkcolor={red!50!black},
	citecolor={blue!50!black},
	urlcolor={blue!80!black}
}
\usepackage[flushleft]{threeparttable}
\usepackage{natbib}
\bibpunct{(}{)}{;}{a}{}{,} 
\usepackage{txfonts}
\usepackage{amsmath}
\usepackage{array}
\usepackage{tabularx} 
\begin{document}

	\title{Distance to the Brick cloud using stellar kinematics }
	
	\author{Á. Martínez-Arranz
		\inst{1}
		\and
		R. Sch\"odel
		\inst{1}
		\and
		F. Nogueras-Lara
			\inst{2}
			\and
		B. Shahzamanian
		\inst{1}
	}
	
	\institute{Instituto de Astrofísica de Andalucía (CSIC), University of Granada,
		Glorieta de la astronomía s/n, 18008 Granada, Spain\\
		\email{amartinez@iaa.es}
		\and
		Max-Planck Institute for Astronomy, K\"onigstuhl 17, 69117 Heidelberg, Germany\\
	}
	
	\date{Received 4 February 2022 / Accepted 21 March 2022}
	
	
	\abstract
	{The central molecular zone at the Galactic center is currently being studied intensively to understand how star formation proceeds under the extreme conditions of a galactic nucleus. Knowing the position of molecular clouds along the line of sight toward the Galactic center has had important implications in our understanding of the physics of the gas and star formation in the central molecular zone.
	It was recently claimed that the dense molecular cloud G0.253 + 0.016 (the Brick) has a distance of $\sim$7.20 kpc from the Sun. That would place it outside of the central molecular zone, and therefore of the nuclear stellar disk, but still inside the Bulge. 

	}
	{
		Theoretical considerations as well as observational studies show that stars that belong to the nuclear stellar disk have different kinematics from those that belong to the inner Bulge. Therefore, we aim to constrain the distance to the Brick by studying the proper motions of the stars in the area.
		
	}
	{
		We used ESO HAWK-I/VLT imaging data from epochs 2015 and 2019 to compute proper motions on the Brick and in a nearby comparison field free of dark clouds. }
	%
	{
		The stellar population seen toward the nuclear stellar disk shows the following three kinematic components: 1) Bulge stars with an isotropic velocity dispersion of $\sim$3.5 micro-arc second per year; 2) eastward moving stars on the near side of the nuclear stellar disk; and 3) westward moving stars on the far side of the nuclear stellar disk. We clearly see all three components toward the comparison field. However, toward the Brick, which blocks the light from stars behind it, we can only see kinematic components 1) and 2).  }
	%
	{
		While the Brick blocks the light from  stars on the far side of the nuclear stellar disk, the detection of a significant component of eastward streaming stars implies that the Brick must be located inside the nuclear stellar disk and, therefore, that it forms part of the central molecular zone. 
	}
	
	
	\maketitle
	%
	
	\section{Introduction}
	 In the Galactic center, at a distance of $\sim$\,8\,kpc \citep{2000,Ghez2008,Genzel2010}, we can find the central molecular zone (CMZ), a flattened structure of gas and dust within a radius of\,$\sim$200\,pc \citep{Kruijssen2014,Tress2020}. Occupying a similar space as the CMZ, there is the so-called nuclear stellar disk (NSD), a rotating structure (\citealp{2015,Shahzamanian2021}), with a radius of $\sim150$\,pc and a scale height of $\sim45$\,pc \citep{Launhardt2002,refId0,Sormani2020_revised}. 
	
	Although most stellar mass in the NSD probably formed\,$\gtrsim$\,8Gyr ago, it has been one of the most active star forming regions of the Milky Way in the past~100 Myr \citep{Matsunaga2011,NoguerasLara2019b}. Nevertheless, the current star formation rate appears to be  about a factor of ten lower than expected, considering the high gas density and dense molecular clouds  present in the area \citep{Longmore2013_revised}. 
	
	In order to fully understand the star formation history in the Galactic center and the evolution of the Galactic Bulge, it is crucial to better understand the structures of the NSD and CMZ (\citealp{Sormani2021}). 
	One important question to address that would help us move toward a better understanding of this region is where the dense molecular clouds are located along the line of sight (\citealp{Kruijssen2015}).
	
	In a recent publication by \cite{Zoccali2021}, the  molecular cloud G0.53+0.016 (the Brick) was suggested to be at a distance of 7.2\,$\pm$\,0.20\,kpc from the Sun, that is to say outside the NSD, but still inside the Bulge. This result is based on the photometric analysis of the Brick stars, which they found brighter than they should be if they were placed at the Galactic center.
	
	This claim was recently challenged by \cite{ANoguerasLara2021}, who found that the Brick presents a stellar population compatible with that of the NSD \citep{NoguerasLara2019b}.\ They concluded that it is located at a distance of 8.4\,$\pm$\,0.5 kpc, which would place it inside of the CMZ and NSD. 
	
	In this Letter we cross-check the work of \cite{ANoguerasLara2021} with kinematic data, considering one of these two possibilities based on the assumption that the Brick is mostly opaque to near-infrared light \citep{Longmore2012} and, therefore, we can only detect stars in front of the cloud.\ The first option being that the Brick is outside of the NSD and inside the Bulge, in which case we would expect to find only stars with typical Bulge proper motions, in other words zero mean proper motion and isotropic velocity dispersion \citep{Kunder,Shahzamanian2021}. Alternatively, the second option being that the Brick is inside the NSD, in which case we would expect to find two different kinematic groups: one belonging to the Bulge population and the other to the NSD itself, that is to say with stars on the near side of the disk moving eastward and those on the far side moving westward \citep{2015, Shahzamanian2021}.

	\section{Data and methods}
	\begin{figure}
		
		\includegraphics[width=\hsize]{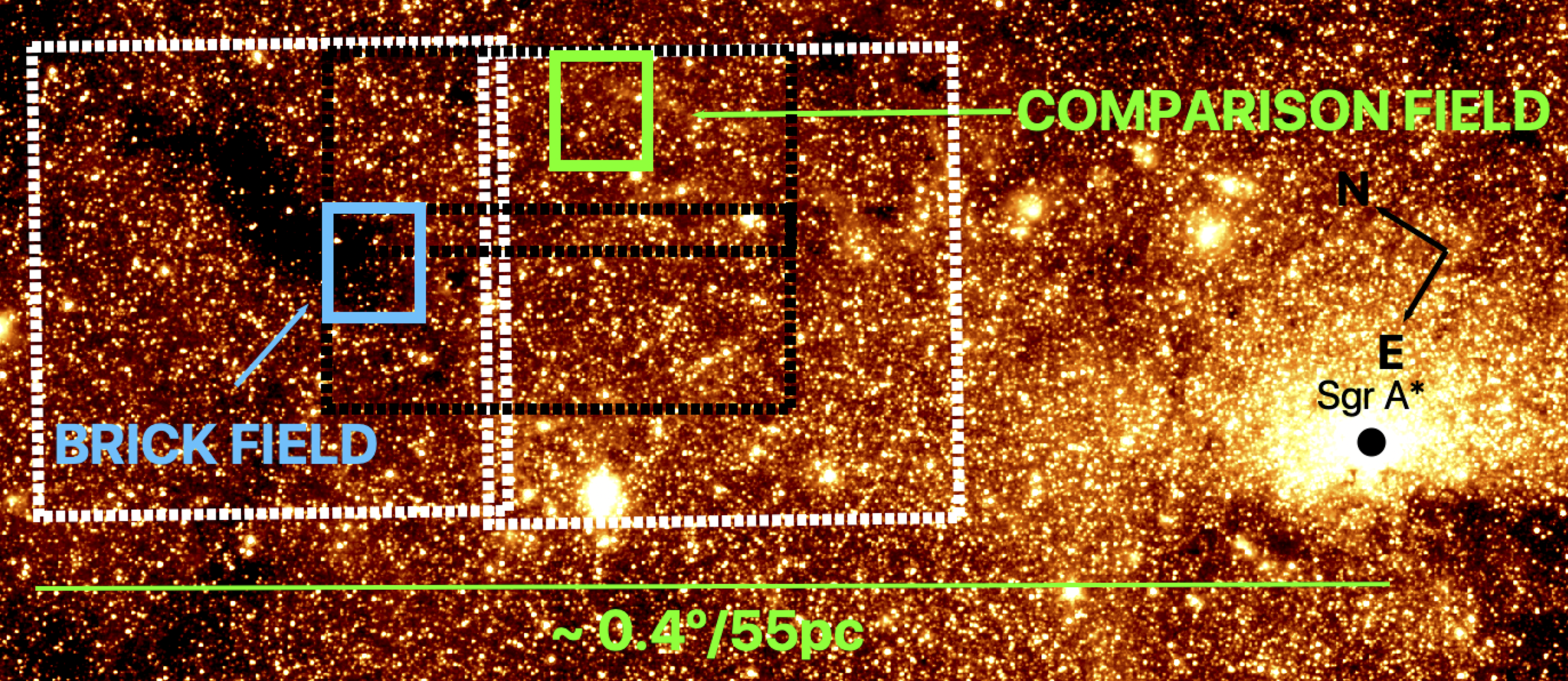}
		\caption{ Pointings from GNS (outlined in black) and D19 (in white) overlaid on a 4.5\,$\mu m $ Spitzer/IRAC image \citep{Sch_del_2014}. The GNS pointings cover smaller areas because of detector windowing. The overlapping areas used for proper motion calculations are shown in blue (on the Brick) and green (comparison field).
		}
		\label{region}
	\end{figure}
	
	 We combined imaging data from the wide-field near-infrared camera HAWK-I/VLT from two epochs, 2015 and 2019\footnote{Based on observations made with ESO Telescopes at the La Silla Paranal Observatory under programmes ID 195.B-0283 and ID 0103.B-0262}. The 2015 data, henceforth D15, correspond to the GALACTICNUCLEUS survey (GNS, \citealp{NoguerasLara2018,NoguerasLara2019a}), acquired with fast photometry mode and reduced with the speckle holography algorithm \citep{10.1093/mnras/sts420} to provide a 0.2" homogeneous angular resolution \citep{NoguerasLara2019a}. The 2019 data, henceforth D19, were acquired with the new ground layer adaptive optics assisted by laser (GRAAL), with an average angular resolution of 0.4" \citep{Zoccali2021} for $H$ and $Ks$  bands.

   The images from both data sets overlap on a small region on the Brick, which is outlined in Fig.\,\ref{region}, along with the comparison field. Both fields have the same dimensions of 0.027º by 0.034º (3.7\,pc by 4.7\,pc), centered on the coordinates shown in Table\,\ref{coordinates}, at a projected distance of $\sim$\,0.22º\,/\,30\,pc and $\sim$\,0.28º\,/\,39\,pc, respectively, to the Galactic east of Sgr A*.
    \begin{table}
     \caption{Comparison and Brick field coordinates}
     \label{coordinates}
     \begin{center}
    		\begin{threeparttable}
     \centering	
    
     \begin{tabular}{ c c c}
     	\hline
     	Field & l(º$: ': ''$) & b(º$: ': ''$) \\
     	\hline
     	Comparison &+0:09:50.946 &+0:02:53.924  \\
     	Brick  & +0:13:43.943 &+0:00:17.924 \\
     	\hline
      \end{tabular}
  	\end{threeparttable}
    \end{center}
    \end{table}

	\subsection{D15 }
	
	The acquisition and reduction of D15 is described in \cite{NoguerasLara2018,NoguerasLara2019a}. Due to saturation issues with the Ks band for stars brighter than $\sim$11.5\,mag, we used only H band data for the calculation of proper motions.\\

	\subsection{ D19}

\begin{figure}[h]
		
		\includegraphics[width=\hsize]{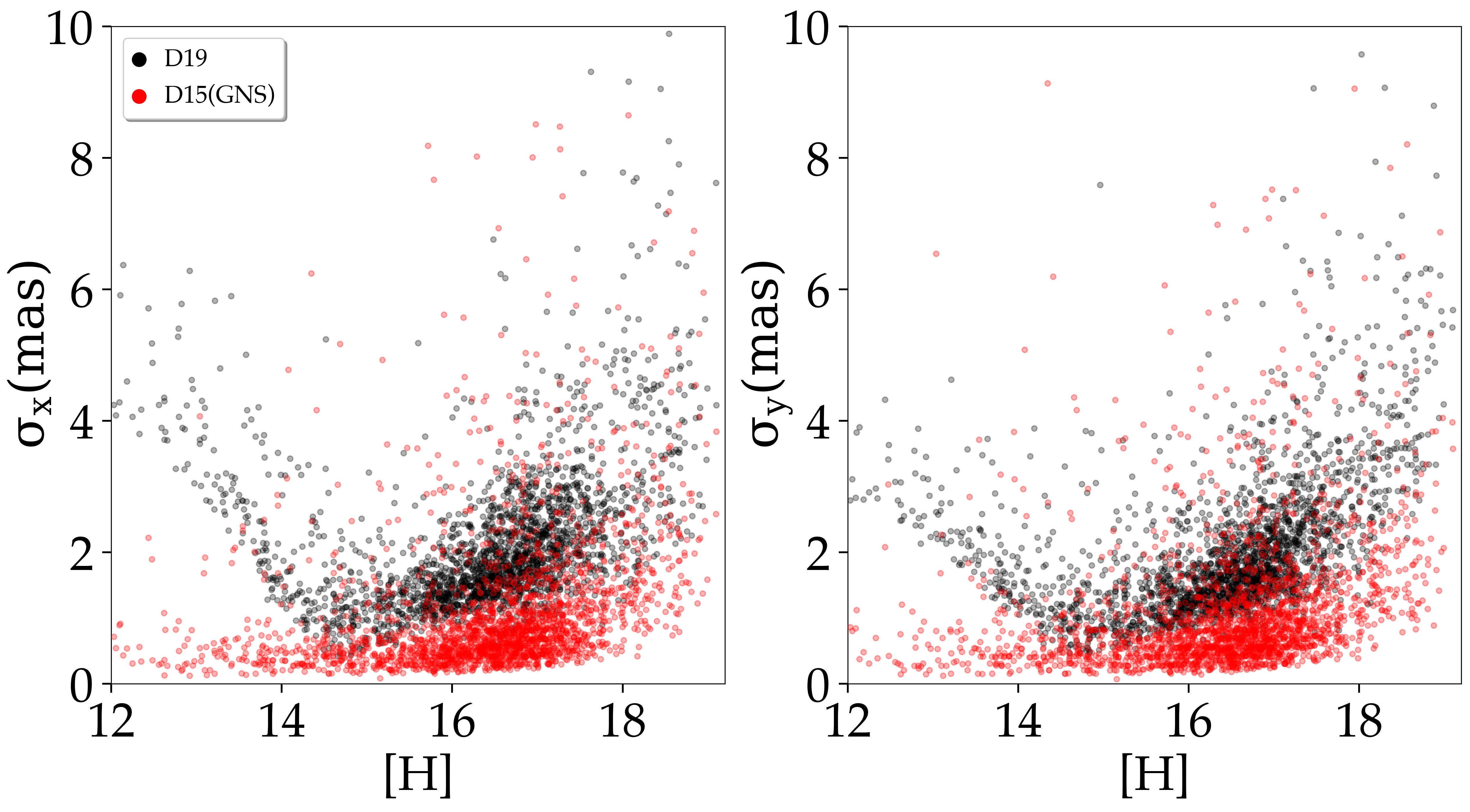}
		\caption{ Relative astrometric uncertainty as a function of H magnitude for all stars in the Brick and comparison fields that are common to D15(red dots) and D19 (black dots). }
		\label{pilot_study}
	\end{figure}

	We used two pointings of D19 (outlined in white in Fig.\ref{region}). These data consist of 15 jittered exposures (DIT\,=\,10s, NDIT\,=\,1) and 32 jittered exposures (DIT\,=\,10s, NDIT\,=\,3), respectively. We applied standard data reduction (sky subtraction, flat fielding, and bad pixel correction). The sky was produced from the jittered images by stacking the different frames  and selecting the pixels with the smallest values. This procedure was necessary because the extreme source density toward the target does not result in a homogeneous sky when applying a median. For the photometry and astrometry, we used \textit{StarFinder} \citep{Diolaiti2000}. In order to compute the mean position and uncertainties of the detected stars, we aligned all jittered frames employing the first one as a reference, using a  linear polynomial and considered stars in different frames to be the same ones if they were less than 1 pixel apart.   
	Due to the long exposure time  of these images, nonlinearity and saturation effects become increasingly important for stars brighter than $H\,=\,14$, resulting in increased astrometric uncertainties (Fig.\,\ref{pilot_study}). Therefore, for the fine alignment of the frames, we considered only stars with magnitudes 14\,$\leq$\,$H$\,$\leq$\,16, which have the smallest position uncertainties according  to Fig.\,\ref{pilot_study}.  We only accepted stars detected in all frames. We used the mean and the uncertainty of the mean of the positions in all frames. We only used areas covered by at least four pointings. We calibrated D19 photometrically via common stars with the SIRIUS/IRSF catalog of \cite{Nishiyama2006}, as was done for the GNS data \citep{NoguerasLara2019a}.

	\subsection{Proper motions}
	We aligned D19 with D15 using common stars. We applied a color cut ${H-Ks\,>\,1.3}$ to remove foreground stars (see \citealp{NoguerasLara2019b}). For an estimation of the displacement and rotation angle between both data sets, we made a first alignment by finding similar three-point asterisms with the \textit{astroalign} package in python \citep{BEROIZ2020100384} and a second one with a polynomial fit (\textit{IDL polywarp}), considering two stars to be the same if they were less than 1 pixel apart. This procedure was iterated until the number of common stars remained stable. We computed the alignment uncertainties using a Jackknife resampling approach; we found that a polynomial of degree\,2 resulted in the most accurate solution.
	
\begin{figure}
		
		\includegraphics[width=\hsize]{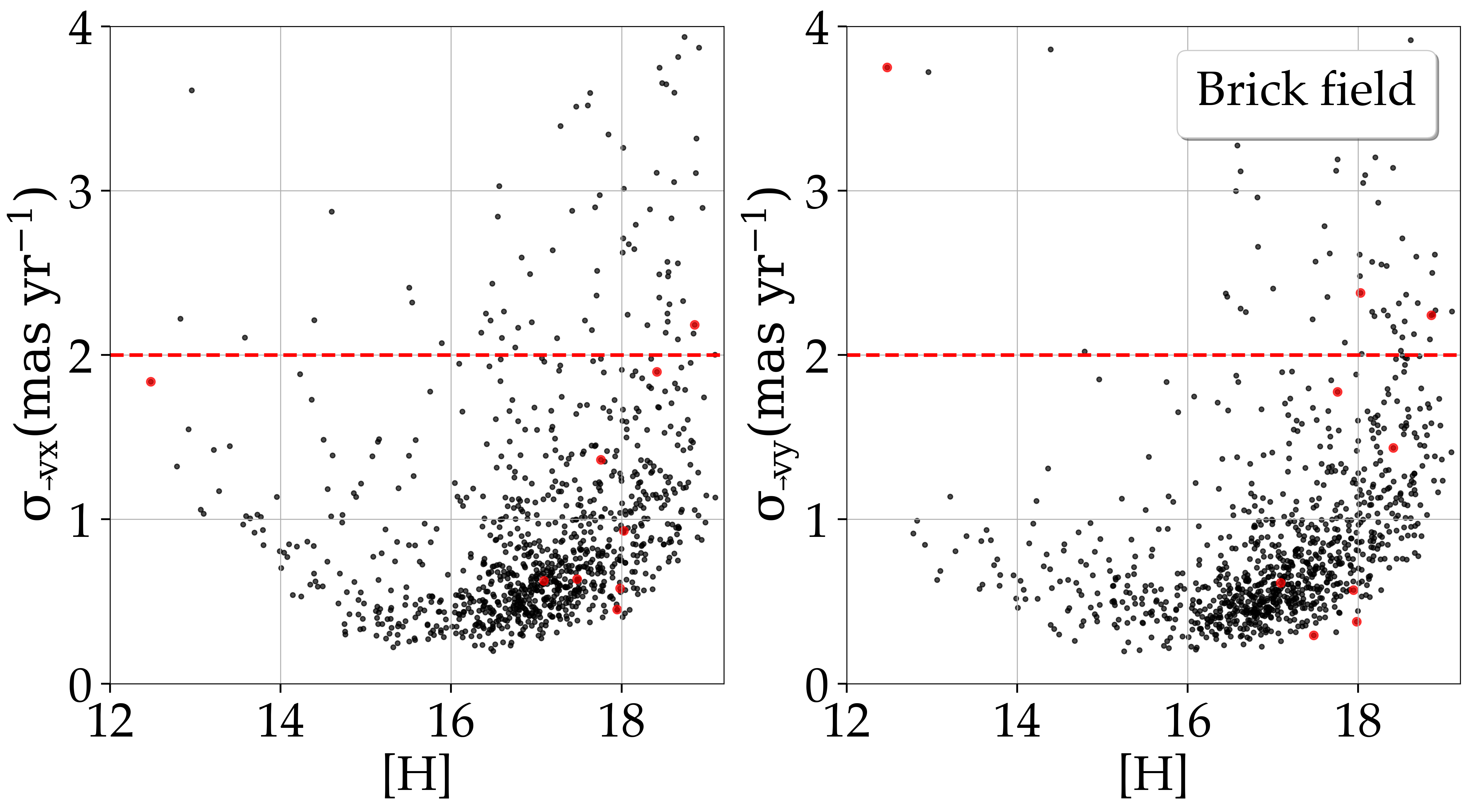}
		 \caption{Uncertainties for $v_{x}$ and $v_{y}$ as a function of H magnitude for stars with colors $H-Ks>1.3$ on the Brick field. We considered only stars with velocity uncertainties smaller than 2\,$mas\,yr^{-1}$\,(red dashed line). We did not consider stars with an absolute difference in magnitude with GNS bigger than 0.5 (red dots).  We applied the same criteria in both fields.
		}
		\label{dvx_vs_mag_ZONES}
	\end{figure}

	After aligning the lists, we computed the velocities for each star by subtracting the positions of common stars and divided them by the time baseline ($\sim$\,4\,year). We computed the uncertainties quadratically, propagating the errors in position for each star and the alignment uncertainty for the D19 stars. From that point on, we considered only stars with proper motion uncertainties below 2.0\,$mas\,yr^{-1}$ and with an absolute magnitude difference with GNS smaller than 0.5 magnitudes (Fig. \ref{dvx_vs_mag_ZONES}).
	
	\section{Results}
We studied stellar kinematics in two fields:  the Brick field and the Comparison field to Galactic west of the Brick on the NSD (Fig.\,\ref{region}).
	
	\subsection{Comparison field}
	\label{Comparison field}
\begin{figure}
	
	\includegraphics[width=\hsize]{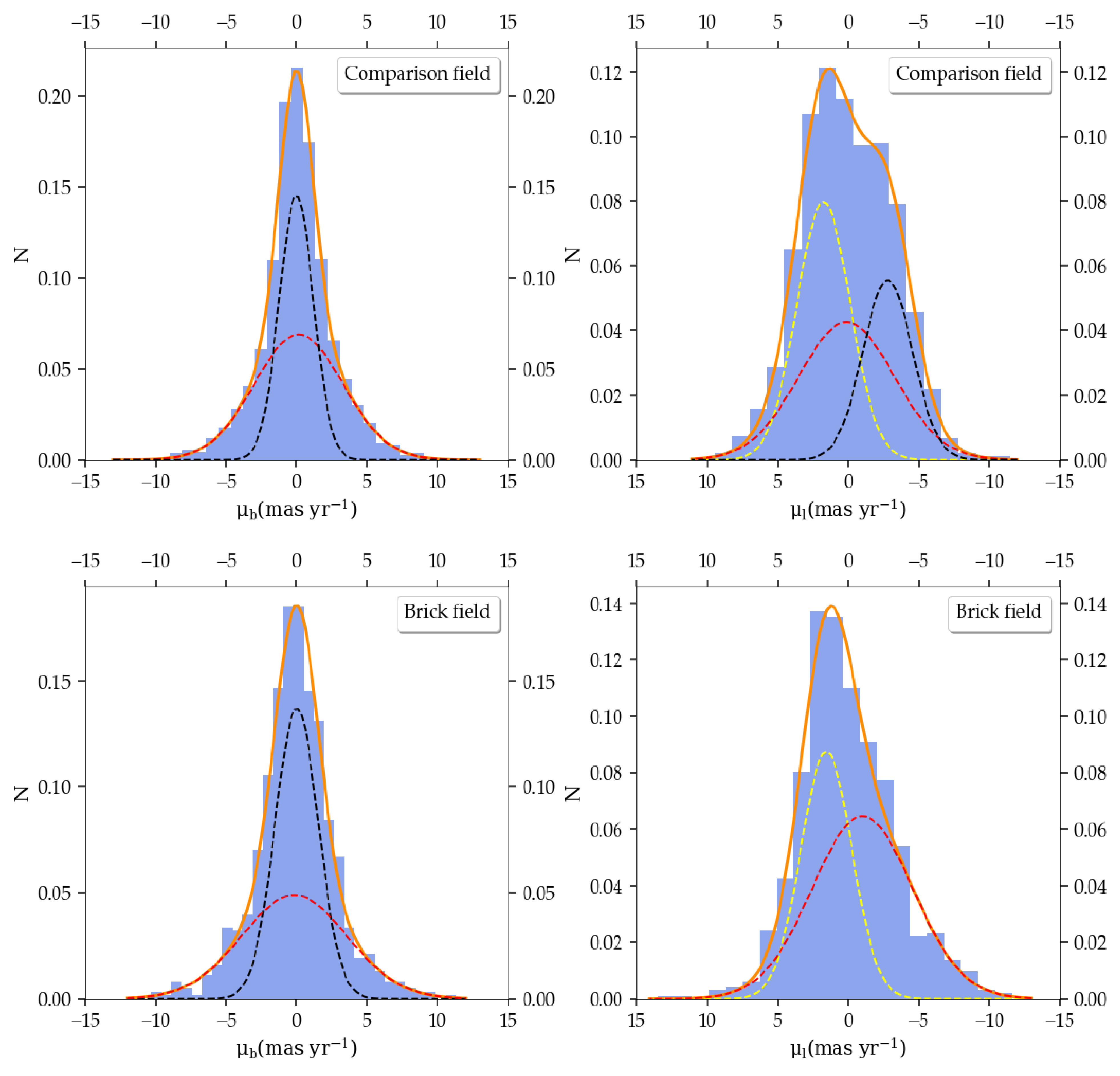}
	\caption{Proper motion distribution for the comparison field (top row) and the Brick field (bottom row). Left-hand plots correspond to the perpendicular component, where the red Gaussian represents the Bulge stars and the black one corresponds to the NSD stars. Right-hand plots correspond to the parallel component, where the red Gaussian represents the Bulge stars and the yellow and black ones represent the NSD stars moving eastward and westward, respectively.  
	}
	\label{all_in_one}
\end{figure}

	 Figure\,\ref{all_in_one}\,(top\,left) shows the proper motion distribution perpendicular to the Galactic plane. We computed the binning of the data  with the python function \textit{ numpy.histogram\_bin\_edges} \citep{Harris2020}, and considered the uncertainty for the height of each bin as the square root of the number of the stars in that bin. We fit it with different Gaussian models, using the package  \textit{dynesty} (\citealp{2020}) in python, which estimates Bayesian posterior probabilities and evidences. We tried doing this with one, two, and three-Gaussians models. To estimate the goodness of each fit, we used the logarithm of posterior probability, $log\,Z$ (see \citealp{Shahzamanian2021}). We found that the best fit is achieved with two Gaussians. 
	We show the best fit Gaussian model along with the proper motion distribution in Fig.\,\ref{all_in_one} (top left) and their parameters in Tab.\,\ref{table:out of brick}; uncertainties correspond to a 1$\sigma$ spread of each parameter's posterior distribution. We interpret the red Gaussian in  Fig.\,\ref{all_in_one} (top left), with the broader velocity dispersion, to represent the stars from the Bulge, in agreement with \cite{Kunder} and \cite{Shahzamanian2021},  and the narrow ones to represent the stars from NSD.

	The proper motions parallel to the Galactic plane show a broader distribution than the ones perpendicular to the Galactic plane (Fig.\,\ref{all_in_one}\,top\,right), which is in agreement with \citealp{Shahzamanian2021}. We followed the procedure explained in the previous paragraph with perpendicular components. We found the best fit corresponding to a three-Gaussians model. The mean Gaussians and the proper motion distribution are shown in Fig. \ref{all_in_one} (top right)  and their parameters are found in Table \ref{table:out of brick}. We interpret the three-Gaussians model in Fig.\,\ref{all_in_one}\, (top\,right) as follows: bulge stars, with about zero proper motion and high velocity dispersion (\citealp{Soto,Shahzamanian2021}); and eastward and westward moving stars on the near and far side of the NSD \citep{2015,Shahzamanian2021}.

\begin{figure}
		
		\includegraphics[width=\hsize]{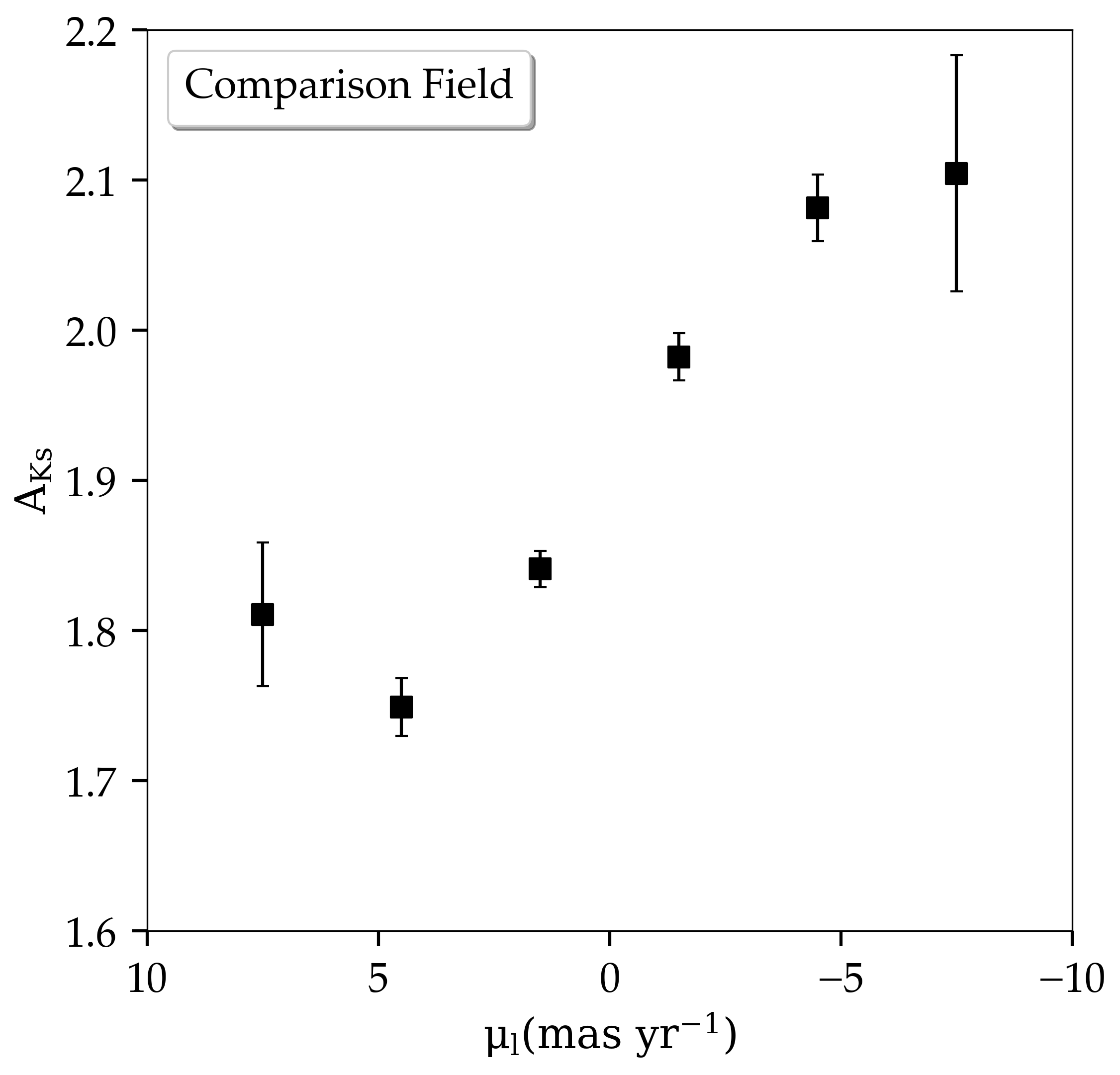}
		\caption{Extinction as a function of $\mu_{l}$ for the comparison field. Squares and bars represent the mean extinctions and their statistical uncertainties for stars in bins of 3\,$mas\ yr^{-1}$ width. Statistical uncertainties for $\mu_{l}$ are smaller than the width of the squares.}
		\label{extintion}
	\end{figure}

	The lower amplitude of the Gaussian, representing the westward moving stars, is probably a consequence of these stars experiencing higher extinction than the eastward moving stars since the former probably belong to the far side of the NSD. To check this, we de-reddened the stars in the comparison field using a star-by-star approach assuming that the intrinsic color $H-Ks$ is similar for all the stars analyzed (see \citealp{BNoguerasLara2021}). In Fig.\ref{extintion}  we show mean Ks extinction as a function of $\mu_{l}$. On average, the extinction of eastward moving stars is smaller than for the westward moving stars, which justifies why we detected fewer westward moving stars (see also Fig.\,10 of \citealp{Shahzamanian2021}).

	To compute the mean proper motion values of the NSD and Bulges stars, since we do not consider an absolute frame of reference, our basic assumption is that the mean proper motion of all stars is zero. This assumption does not appear to be valid in the comparison field since we detected more stars from the near side of the NSD than from the far side, that is, there is a net eastward proper motion present in the field. This makes our reference frame drift eastward and this results in an underestimation of eastward motion and an overestimation of westward motions.
	If we correct this by assuming the same mean eastward and westward velocities, then we see that the near and far sides of the NSD rotate with approximately 2.27\,$mas\,yr^{-1}$ which, within uncertainties, agrees with measurements and theoretical expectations (\citealp{Shahzamanian2021,2015,Sormani2021}). The mean velocities ($\overline\mu_{b}$ and $\overline\mu_{l}$)  and the velocities' dispersion ($\sigma_{b}$ and $\sigma_{l}$ ) for the Bulge population  are in agreement with the values found in the literature \citep{Soto,Kunder,Shahzamanian2021} and also theoretical studies (\citealp{Sormani2021}).

	\begin{table}
		\begin{center}
			\begin{threeparttable}

					\caption{Best-fit parameters and uncertainties for the comparison field data (Fig.\,\ref{region})}   
				\label{table:out of brick}      
				\centering                          
				\begin{tabular*}{\columnwidth}{c|@{\extracolsep{\fill}}c c c c }        
					\hline \multicolumn{4}{c}{Comparison field }\\
					\hline  \hline               
					$\textit{perpendicular}$ &Bulge& NSD&-\\    
					\hline                     
					
					${\mu_{b}}\  (mas\ yr^{-1})$        &$0.15^{+0.13}_{-0.12}$    &  -0.02$\pm$0.08& -\\             
					${\sigma_{b}}\  (mas\ yr^{-1})$ &  3.14$\pm$0.16    &$1.21^{+0.10}_{-0.11}$&-\\             
					${amp_{b}}$                           &0.54$\pm$0.06   & 0.44$\pm$0.06&-\\            
					\hline\hline                                  
					$\textit{parallel}$&Bulge& NSD&NSD\\
					\hline                     
					${\mu_{l}}\  (mas\ yr^{-1})$          & $0.10^{+0.50}_{-0.49}$  &$-2.77^{+0.43}_{-0.45}$  &$1.76^{+0.37}_{-0.35}$\\                                                   
					${\sigma_{l}}\  (mas\ yr^{-1})$      &$3.39^{+0.26}_{-0.24}$  &$1.78^{+0.30}_{-0.29}$  &$1.90^{+0.30}_{-0.31}$\\                                    
					${amp_{l}}$                                 &$0.35^{+0.13}_{-0.12}$ & $0.25^{+0.08}_{-0.09}$  &$0.38^{+0.11}_{-0.12}$\\                                         
					\hline\hline 
				\end{tabular*}
				\begin{tablenotes}
					\small
				
					\item\textbf{Notes:} ${\mu}$, ${\sigma}$, and  ${amp}$ are the mean velocity, standard deviation, and amplitude of the Gaussians fitted to the distribution. Uncertainties correspond to a 1$\sigma$ spread of each parameter's posterior distribution. {Perpendicular} and {parallel} indicate the proper motions' direction in relation to the Galactic plane.  
				\end{tablenotes}
			\end{threeparttable}
		\end{center}
	\end{table}

	\subsection{Brick field}

	In the Brick field, we proceeded as we did in the comparison field for the perpendicular component. Also, in this case, two Gaussians optimally describe the data (Fig.\,\ref{all_in_one}, bottom left and Table \ref{table:on brick}). These results agree with the values found for the comparison field for the perpendicular components.
	
	For the parallel component, the three-Gaussians model does not converge to make a single solution. The best fit to the distribution corresponds to a two-Gaussians model (Fig.\,\ref{all_in_one}\,bottom\,row and Table\,\ref{table:on brick}). The absence of a third Gaussian in the model could mean that the mean opacity on the Brick is  too high to observe stars through it, blocking the light from the stars in the far side of the NSD, which agrees with the results found by \cite{ANoguerasLara2021}. This is not surprising because the Brick is an extremely dense mid-infrared dark cloud (\citealp{Longmore2012}). Another possibility for the presence of only two Gaussians is that the number of detected sources through the Brick is not large enough to affect the distribution significantly.
	
	As in the comparison field, in this case, with no sources moving westward (or just a small number of them),  the reference frame drifted eastward, and in a more pronounced way because of the extreme extinction caused by the Brick. As a consequence, the mean velocity for the Bulge Gaussian is not around zero any more and the velocity of eastward moving stars is underestimated. Since we know that the mean velocity of Bulge stars should be around zero, we can compensate for the drifting effect by requiring the mean motion of the Bulge stars to be zero. That leaves us with a mean velocity for the parallel component of the NSD  of $\mu_{east}=1.57+0.84= 2.41\  mas\ yr^{-1}$. This value  agrees, within the uncertainties, with our results in the comparison field and with the rotation velocities for the disk found by \cite{Shahzamanian2021} and \cite{2015}.\\

	\begin{table}
		\begin{center}

			\begin{threeparttable}
				\caption{Best-fit parameters and uncertainties for the Brick field data  (Fig.\,\ref{region})}      
				\label{table:on brick}      
				\centering                          
				
				\begin{tabular*}{\columnwidth}{c|@{\extracolsep{\fill}}c c   }        
					\hline \multicolumn{3}{c}{Brick field }\\
					\hline  \hline               
					$perpendicular$&Bulge& NSD\\
					\hline                     
					${\mu_{b}}\  (mas\ yr^{-1})$  &	$-0.16^{+0.25}_{-0.24}$	      &  $-0.02^{+0.12}_{-0.11}$	\\                                                   
					${\sigma_{b}}\  (mas\ yr^{-1})$&$3.73^{+0.18}_{-0.19}$    &  1.53$\pm$0.12\\                                    
					${amp_{b}}$  &0.45$\pm$0.06      &  0.53$\pm$0.06\\                
					\hline            
					\hline                   
					$parallel$ &Bulge& NSD\\    
					\hline               
					${\mu_{l}}\  (mas\ yr^{-1})$        & $-0.84^{+0.51}_{-0.52}$ &$1.57^{+0.33}_{-0.31}$\\             
					${\sigma_{l}}\  (mas\ yr^{-1})$ &$3.44^{+0.19}_{-0.21}$  &$1.81^{+0.40}_{-0.43}$\\             
					${amp_{l}}$          & $0.59^{+0.14}_{-0.15}$  &$0.38^{+0.16}_{-0.15}$\\            
					\hline                               
					\hline                     

				\end{tabular*}
				
				\begin{tablenotes}
					\small
					\item\textbf{Notes:} ${\mu}$, ${\sigma}$, and  ${amp}$ are the mean velocity, standard deviation, and amplitude of the Gaussians fitted to the distribution. Uncertainties correspond to a 1$\sigma$ spread of each parameter's posterior distribution. {Perpendicular} and {parallel} indicate the proper motions' direction in relation to the Galactic plane.
				\end{tablenotes}
			\end{threeparttable}
		\end{center}
	\end{table}

	\section{Conclusions}
	We analyzed the stellar kinematics for two different regions, one on the dense molecular cloud G0.253+0.016 (the Brick) and another on a Comparison field westward from it, delimited with blue and green in Fig. \ref{region}.
	We found that, in both cases, the models that best describe the parallel and perpendicular components of the proper motion distributions are compatible with the presence of two different populations.

   In the comparison field, for the perpendicular component of the velocity, we can optimally fit the proper motion distribution with two Gaussians (Fig. \ref{all_in_one}\,top\,left), representing a Bulge and NSD population. We interpret the broader Gaussian to characterize the stars from the Bulge and the narrow one to correspond to the population of the NSD. Three Gaussians optimally describe the proper motion distribution for the parallel component: with one corresponding to the Bulge population and the other two corresponding to the NSD population, moving eastward and westward, respectively (Fig.\,\ref{all_in_one}\,top\,right).
In the Brick field, as for the comparison field, we clearly identify two different populations, which are represented with two Gaussians for the perpendicular component (Fig.\,\ref{all_in_one}\,bottom\,left) and, because of extinction, also two Gaussians for the parallel component (Fig.\,\ref{all_in_one}\,bottom\,right).

	The velocity dispersion values for the Bulge populations in both fields are in agreement with the ones found by \cite{Soto}, \cite{Kunder}, and \cite{Shahzamanian2021}. With respect to the parallel component, the values obtained in both fields are compatible with each other and also with the rotation velocity for the NSD found by \cite{2015} and \cite{Shahzamanian2021}.
	
	 If the Brick were located at a distance from the Sun of $\sim$\,$7.20$\,kpc, as claimed by \cite{Zoccali2021}, it would be placed well outside the NSD, but still inside the Bulge. Consequently, we would not expect to find traces of two different populations in the distribution, but only of one. Instead, we can clearly distinguish two different populations.\ Furthermore, the parameters of these distributions are in agreement with those found by other authors for the NSD and Bulge populations. These are strong indicators suggesting that the Brick is embedded in the NSD and the CMZ, and hence that it is located at the Galactic center. 
	 
	This conclusion is supported by different arguments in other studies. For example, the turbulent velocity dispersion found for the gas in the Brick cloud does not match that found in other molecular clouds in the Galactic plane, which are significantly lower (see e.g., \citealp{Henshaw2016,Henshaw2019}).

	\begin{acknowledgements}
		      We acknowledge financial support from the State Agency for Research of the Spanish MCIU through the ``Center of Excellence Severo Ochoa'' award for the Instituto de Astrof\'isica de Andaluc\'ia (SEV-2017-0709).\\
      We acknowledge financial support from national project PGC2018-095049-B-C21 (MCIU/AEI/FEDER, UE).\\
      FN-L gratefully acknowledges support by the Deutsche Forschungsgemeinschaft (DFG, German Research Foundation) – Project-ID 138713538 – SFB 881 (“The Milky Way System”, subproject B8), and the sponsorship provided by the Federal Ministry for Education and Research of Germany through the Alexander von Humboldt Foundation. \\

	\end{acknowledgements}
	
	\bibliographystyle{aa} 
	
	\bibliography{lib_manual_new.bib} 

\end{document}